# Self-assembling T7 phage syringes with modular genomes for targeted delivery of penicillin against β-lactam-resistant *Escherichia coli*

Hyunjin Shim[1,*]

## Author Information

### Affiliations

[1]Department of Biology, California State University, Fresno, 5241 N Maple Ave, Fresno, CA 93740, USA

*Corresponding author: Hyunjin Shim (jinenstar@gmail.com)

### Orcid links

Hyunjin Shim orcid=0000-0002-7052-0971

### Keywords

Antimicrobial resistance; Antibiotics; Phage therapy; Phage reboot; β lactam; Penicillin

# Abstract

Bacteriophages are promising alternative antimicrobial agents due to their high specificity for host bacteria and minimal immunogenicity in humans. However, their therapeutic application is limited by their nature as biological entities, which can lead to unintended evolutionary consequences such as horizontal gene transfer. In this study, we address these challenges by repurposing only the structural components of bacteriophages as vesicles to deliver antibiotics directly into the cytoplasm of bacterial hosts. This approach is based on two key hypotheses: first, antibiotics such as β-lactams remain effective against resistant bacteria if injected directly into the cytoplasm, bypassing resistance mechanisms; second, phage structures can be synthesized and self-assembled *in vitro* using modular genomes and cell-free protein expression systems to carry small molecules such as antibiotics as cargo. To test these hypotheses, we utilized T7 phages and penicillin-resistant *Escherichia coli* as a model system. First, we designed the T7 phage genome into a modular format containing only the genes encoding structural components and synthesized the gene fragments via de novo gene synthesis. These phage structures were then rebooted *in vitro* using a cell-free protein expression system in the presence of penicillin G, allowing the antibiotics to be incorporated as cargo during the self-assembly process. Finally, we tested the antimicrobial activity of these antibiotic-loaded phage syringes against penicillin-resistant *E. coli*. The results demonstrate that phage syringes effectively reduce the population of penicillin-resistant *E. coli* compared to negative controls, including free penicillin and water. However, their efficacy was lower than that of the positive control, natural T7 phages. This study highlights the potential of using phage structures as antibiotic delivery vehicles, offering a novel strategy to overcome both antibiotic resistance and the limitations of phage therapy.

# Introduction

Bacteria have significantly influenced human health in diverse ways. On the positive side, bacteria can play a beneficial or commensal role, such as probiotic bacteria within the gut microbiome, which support digestion, immune function, and overall health [1–3]. Conversely, bacteria have also been the cause of devastating infectious diseases throughout history. Pathogenic bacterial outbreaks, such as the bubonic plague or cholera epidemics, have not only caused widespread mortality but also reshaped the course of civilizations, leaving lasting impacts on societies worldwide [4]. Since the discovery of penicillin, humanity has greatly benefited from a range of antibiotics derived from microorganisms such as fungi and soil bacteria [5,6]. During the golden age of antibiotics, it appeared that pathogenic bacteria would no longer significantly influence human development [7]. However, bacteria, having co-evolved with their hosts for millions of years, possess diverse evolutionary strategies to overcome challenges posed by antimicrobial agents [8–10].

The adaptation of bacteria to the β-lactam class of antibiotics exemplifies their ability to develop resistance through multiple mechanisms [8]. β-lactam antibiotics, including penicillin, are small molecules that kill bacteria by inhibiting cell wall synthesis, an essential process for bacterial survival [11]. For instance, penicillin G enters bacterial cells through porins in the outer membrane of gram-negative bacteria and disrupts cell wall synthesis by binding to critical cell wall components [12]. Resistance to penicillin G was first observed in the early 1940s, even before antibiotics were introduced to the market [13]. The primary resistance mechanisms to β-lactams differ between bacterial groups due to differences in cell envelope structure: gram-positives modify their penicillin-binding proteins (PBPs), while gram-negatives rely on β-lactamase production and outer membrane control of antibiotic entry [8]. Furthermore, some bacteria reduce antibiotic influx by reducing porin expression, while others deploy efflux pumps to actively expel antibiotics [14]. Multidrug-resistant bacteria exhibit resistance to multiple classes of antibiotics through various biochemical pathways, with gram-negative bacteria such as *Escherichia coli* increasingly becoming resistant to almost all currently available antibiotic classes [15].

The rise of antibiotic resistance against β-lactam antibiotics prompted the scientific community to develop new variants and alternative classes of antibiotics with diverse mechanisms of action [16]. These new drugs targeted DNA synthesis, RNA synthesis, protein synthesis, and metabolic pathways [9,17]. Despite these advancements, bacteria have continued to evolve resistance to each new class of antibiotics [18,19]. By the 2000s, reports of superbugs resistant to all known antibiotics began to surface, raising alarms within the scientific and healthcare communities [18]. This resurgence of untreatable bacterial infections has brought humanity closer back to a pre-antibiotic era, where minor surgical procedures or routine infections could become life-threatening due to nosocomial infections in hospital settings [19].

The spread of antimicrobial resistance is fueled by multiple factors tied to the current practices and economics of antibiotic use [5,19]. In response to the growing threat of superbug infections, the scientific community has been exploring alternative antimicrobial strategies beyond small-molecule antibiotics [20–22]. Phage therapy uses bacteriophages - viruses that infect and kill bacteria - to combat antibiotic-resistant pathogens [23]. However, phage therapy involves biological entities capable of replication within the human body, potentially leading to unintended evolutionary consequences [24–27]. As a result, its use has been restricted primarily to compassionate cases involving patients infected with superbugs [28]. While phage therapy shows promise, concerns about biosafety and regulatory challenges remain barriers to its widespread application [29].

In this study, we explore the potential of using bacteriophages as vesicles to deliver antibiotic molecules directly into the cytoplasm of bacteria. We hypothesize that phages can function as carriers of antibiotic molecules based on recent observations that phages can encapsulate cargo beyond their genetic material. For example, a recent study demonstrated that a family of jumbo phages carry nucleus-like compartments that shield phage DNA from destruction by bacterial defense systems [30–32]. Furthermore, we hypothesize that phages can specifically bind to their host bacteria and inject antibiotic molecules into the cytoplasm through their natural infection mechanisms. As a proof-of-concept, we engineered the genomic elements of T7 phages, which are lytic phages of *Escherichia coli*, to reboot in a cell-free protein expression system (Figure 1). During this process, varying doses of penicillin G were introduced as cargo, to be spontaneously incorporated into the phage capsids during their self-assembly. We term these antibiotic-loaded vesicles "phage syringes," as they utilize the phages' specific binding and invasion mechanisms to deliver antibiotics directly into the bacterial cytoplasm.

We validated the assembly and functionality of these phage syringes through several techniques. First, we quantified the phage syringes using protein quantification assays and examined their structural integrity using Transmission Electron Microscopy (TEM). Next, we evaluated the antimicrobial activity of the phage syringes against penicillin-resistant *E. coli* strains using antimicrobial susceptibility testing. To demonstrate the effectiveness of phage syringes, we compared their activity against five controls: standard T7 phages (positive control), antibiotic-free T7 phage syringes with all proteins (positive control) or with only structural proteins(negative control), free penicillin G (negative control), and HPLC-grade water (negative control). This study investigates the feasibility of phages as targeted delivery vehicles for antibiotics, overcoming resistance mechanisms that prevent antibiotics from penetrating the bacterial cell membrane and wall. By leveraging the specificity of phage-bacteria interactions, we aim to demonstrate that these self-assembling phage syringes offer a novel alternative antimicrobial strategy that is both specific and effective.

## Results

### Rebooting T7 phages with modular genomes shows a high protein concentration

We designed the gene fragments of T7 phage genomes into modular segments and rebooted these phages using cell-free protein expression systems (Figure 2). The modular genomes enable the flexibility of phages to be rebooted using all the proteins (PSAP) or only the structure proteins (PSSP). The DNA quality control of these modular phage genomes synthesized de novo showed high concentrations of nucleic acids and acceptable concentrations of contaminants (Table 1). The T7 phage syringes with structure proteins (PSSP) were loaded with antibiotics by rebooting in the presence of 1 μL, 2 μL, and 10 μL of penicillin G at the concentration of 100 mg/mL (labeled as PS1P, PS2P, and PS10P, respectively).

The most common method to measure bacteriophage concentration involves performing a plaque assay, where serial dilutions of a phage sample are added to a bacterial lawn on an agar plate and counting visible plaques of lysed bacteria, expressed as plaque forming units (PFU) [33]. However, as these phage syringes have been rebooted using de novo gene synthesis and cell-free expression in modular genomes, the concentration of these phage syringes was approximated with A280 readings and A260 readings from a spectrophotometer to measure the protein concentration of phage capsids and the impurity concentration of nucleic acids (Table 2). The A280 application quantifies the concentration of proteins that contain amino acids such as tryptophan, tyrosine, and cys-cys disulfide bonds that exhibit absorbance at 280 nm [34].

The A280 readings show that all the rebooted samples, including the positive controls and phage syringes, had high concentrations of proteins (Figure 3). Notably, the sample of phage syringes with all proteins (PSAP) has the highest protein concentration at 697.847 mg/mL, even compared to that of the positive control of standard T7 phages (PC) at 584.702 mg/mL. However, all the samples contained high impurities of nucleic acids with the A260/280 ratio measured to be above 2. As expected, the sample of phage syringes rebooted with 10 μL penicillin G (PS10P) had the lowest protein concentration at 341.580 mg/mL and the highest A260/280 ratio at 3.94. These results indicate that all the samples still contained intact gene fragments from the cell-free protein expression step, and the phage syringes synthesized with high concentrations of antibiotics were the least effective in protein expression.

### TEM images verify intact capsids and small molecules of T7 phage syringes

To determine the morphology, we visualized the T7 phages rebooted through de novo gene synthesis and cell-free protein expression with Transmission Electron Microscopy (TEM). Each sample was negatively stained, and the images of the samples were generated at 100 nm resolution (Figure 4). Overall, the TEM images of all the samples were dense with background residues, potentially the remains of gene fragments and protein products from cell-free protein expressions (Figure 4). Despite these residues, the TEM image of the standard T7 phages (PC) shows several icosahedral head structures with short and noncontractile tails (Figure 4A), which is consistent with the morphology of bacteriophages in the Order of *Caudovirales* [35]. The TEM image of the T7 phage syringes with all proteins (PSAP) also shows several icosahedral head structures, but short and noncontractile tails are less visible (Figure 4B). Both of these samples contained phage particles that were smaller than 100 nm, which is consistent with the internal structure of the T7 phage capsid at 60 nm in diameter [36]. For the T7 phage syringes with 1 µL penicillin (PS1P), the TEM image shows enlarged capsids of the icosahedral head structures (Figure 4C), which are around 100 nm in diameter. Furthermore, the TEM image of the PS1P sample has visible white dots, which can be inferred as the residual small molecules of penicillin G with a typical size of sub-20nm [37].

The TEM images confirm the structural integrity of phage capsids self-assembled through de novo gene synthesis and cell-free protein expression, using the modular genome design and antibiotics additive. Additionally, these images highlight the need for filtering small molecules and proteins prior to downstream experiments to accurately assess lytic activity. Therefore, we filtered the phage samples using size-exclusion chromatography to remove molecules smaller than 2 kDa in size before conducting antimicrobial susceptibility testing.

## Free penicillin G is ineffective against penicillin-resistant *E. coli*

To show the effectiveness of phage syringes, we added rebooted phage samples to the standard bacterial suspension of *Escherichia coli* and monitored the absorbance at OD600 over the course of 240 minutes to estimate the changes in the colony-forming unit (CFU). For this antimicrobial susceptibility test, we used a strain of *Escherichia coli* (ATCC® 25922) known to be resistant against penicillin G [38].

As a negative control to confirm the resistance of this standard bacterial suspension of *E. coli* against penicillin G, we added 10 µL penicillin G in distilled and sterile-filtered water to the bacterial suspension directly after the first measurement and monitored OD600 readings every 20 minutes for 12 time points (Table 3). The initial CFU of the standard bacterial suspension varied between 0.65 to 2.31$\times 10^8$ cells/mL for the negative control with penicillin (Pen10ul) replicates when measured with OD600 readings from a spectrophotometer (Table S1). The changes in OD600 readings show a slow decline in the CFU of the standard bacterial suspension of *E. coli* throughout the 12 time points. At the end of the test, less than 40% of the initial bacterial population remained in all three replicates (Figure 5A). This decline in the bacterial population is similar to the negative control conducted with High Performance Liquid Chromatography (HPLC) water, as described below.

For the negative control with a blank, we added 10 µL HPLC water to the samples directly after the first measurement and monitored OD600 readings every 20 minutes for 12 time points (Table 3). The initial CFU of the standard bacterial suspension varied between 0.92 to 2.34$\times 10^8$ cells for the negative control with HPLC water (NC) replicates when measured with OD600 readings from a spectrophotometer (Table S1). At the end of the test, less than 60% of the initial bacterial population remained in all three replicates (Figure 5B). The result from this negative control is consistent with the spontaneous death rate of the bacterial population after reaching the stationary phase, during which the viable cell population goes through exponential decline due to nutrient depletion and waste accumulation [39].

## Rebooted standard and synthetic T7 phages are effective against penicillin-resistant *E. coli*

To show the effectiveness of these standard phages against their natural host, we challenged the standard bacterial suspension of *Escherichia coli* with two types of positive controls: the first with standard T7 phages and the second with the phage syringes all proteins (PSAP).

For the positive control with standard T7 phages, we added 1 µL of T7 phages (PC) rebooted from the standard DNA into 600 mL of the bacterial suspension directly after the first measurement and monitored the changes in CFU every 20 minutes for 12 time points (Table 3). The initial CFU of the standard bacterial suspension varied between 2.25 to $2.7 \times 10^8$ cells/mL for the positive controls (PC) replicates when measured with OD600 readings from a spectrophotometer (Table S1). At the end of the test, less than 3% of the initial bacterial population remained in all three replicates (Figure 5B). This decline in the bacterial population is the fastest among all the samples, indicating that rebooted standard T7 phages are the most effective against penicillin-resistant *E. coli* (Figure 5).

For the positive control with T7 phage syringes, we added 1 µL of T7 phage syringes with all proteins (PSAP) rebooted from the synthetic modular genomes into 600 mL of the bacterial suspension directly after the first measurement and monitored the changes in CFU every 20 minutes for 12 time points (Table 3). The initial CFU of the standard bacterial suspension varied between 2.67 to $2.71 \times 10^8$ cells/mL for the phage syringes all proteins (PSAP) replicates when measured with OD600 readings from a spectrophotometer (Table S1). At the end of the test, less than 1% of the initial bacterial population remained in all three replicates (Figure 5C). Notably, two of the PSAP replicates had negative or undetectable OD600 readings after 160 minutes at the 8th time point. A negative OD600 value on a spectrophotometer indicates a measurement error, likely due to a sample being too dilute to accurately read. This decline in the bacterial population is slower than that from the standard T7 phages, but the T7 phage syringes with all proteins (PSAP) are more effective in reducing the end population of penicillin-resistant *E. coli* (Figure 5).

## Penicillin-loaded phage syringes are effective against penicillin-resistant *E. coli*

To show the effectiveness of the penicillin-loaded phage syringes against resistant bacteria, we added 1 µL of phage syringe samples into 600 mL of the standard bacterial suspension of *Escherichia coli* and monitored the absorbance at OD600 over the course of 240 minutes to estimate the changes in the colony-forming unit (CFU).

As a negative control for T7 phage syringes, we added 1 µL of T7 phage syringe with structure proteins only (PSSP) rebooted from the modular genomes directly after the first measurement and monitored the changes in CFU every 20 minutes for 12 time points (Table 3). The initial CFU of the standard bacterial suspension varied between 2.7 to $2.71 \times 10^8$ cells/mL for the phage syringes structure proteins (PSSP) replicates when measured with OD600 readings from a spectrophotometer (Table S1). At the end of the test, less than 28% of the initial bacterial population remained in all three replicates (Figure 5). This decline in the bacterial population is faster than the other negative controls conducted with HPLC water (NC) and free penicillin G (Pen10ul), but slower than the positive controls conducted with standard T7 phages (PC) and phage syringes with all proteins (PSAP).

      Next, we challenged the standard bacterial suspension of *E. coli* with the phage syringes loaded with varying doses of penicillin G (PS1P, PS2P, and PS10P). For these samples, we added 1 µL of T7 phage syringes with structure proteins (PSSP) rebooted from the modular genomes directly after the first measurement and monitored the changes in the CFU every 20 minutes for 12 time points (Table 3). The initial CFU of the standard bacterial suspension varied between 2.67 to 2.72× $10^8$ cells/mL for the penicillin-loaded phage syringe samples and replicates when measured with OD600 readings from a spectrophotometer (Table S1). At the end of the test, less than 9%, 8%, and 29% of the initial bacterial population remained in all three replicates, for PS1P, PS2P, and PS10P, respectively (Figure 5). This decline in the bacterial population is faster than all the negative controls conducted with HPLC water (NC), free penicillin G (Pen10ul), and phage syringes with structure proteins (PSSP), but slower than all the positive controls conducted with standard T7 phages (PC) and phage syringes with all proteins (PSAP). This reduced antimicrobial activity compared to the positive controls was particularly pronounced for PS10P, indicating that rebooting phage syringes with a high dose of antibiotics may reduce the efficacy of their antimicrobial activities.

## Discussion

Antibiotics are widely regarded as one of the greatest achievements of modern medicine, having saved countless lives and significantly reduced human suffering caused by bacterial infections [5]. However, our reliance on small molecules as antimicrobial treatments, due to their effectiveness and affordability, has left us with a limited arsenal to combat pathogenic bacteria [19,40]. While small molecules have many advantages, it has been consistently demonstrated that bacteria possess an ability to adapt and develop resistance, even to newly developed drugs [17,41]. As we face an alarming shortage of new classes of small-molecule antibiotics, attention has shifted toward alternative strategies, such as antimicrobial auxiliary agents and phage therapy [42,43]. Bacteriophages, the natural predators of bacteria, are highly specific and show great promise as antimicrobial agents. Moreover, there is minimal evidence suggesting they are immunogenic to humans, especially given the growing recognition that bacteriophages are an integral component of our microbiomes, maintaining microbial community balance [3,44,45]. However, their use as antimicrobial agents raises biosafety concerns due to their ability to evolve through complex host-parasite interactions [29]. In response to these challenges, there have been ingenious efforts to engineer phages and associated components into safer, more predictable biological agents [46–48].

      Here, we present a proof-of-concept study in which we engineer bacteriophages to deliver antibiotics directly into the cytoplasm of their bacterial hosts through modular genome design, de novo gene synthesis, and cell-free protein expression. This approach leverages the natural specificity and injection mechanisms of bacteriophages [49,50], overcoming the challenge of antibiotic penetration into bacterial cytoplasm posed by resistance mechanisms. We term these engineered phages "phage syringes," as they are syringe-like structures loaded with antibiotic cargo that retain contractile injection systems into the bacterial cytoplasm. The primary objective of this study was to test the hypothesis that phage syringes can mechanically inject small-molecule cargo, other than nucleic acids, into bacterial cells. This hypothesis is based on the observation that small molecules under 20 nm in size [37] are smaller than most phage genomes and may be incorporated as cargo during the spontaneous self-assembly of phage capsids, provided that sufficient concentrations of antibiotics are present during *in vitro* expression. To test this hypothesis, we selected T7 phages as the delivery vehicle, penicillin as the antibiotic cargo, and penicillin-resistant *Escherichia coli* as the bacterial host.

      To demonstrate the efficacy of penicillin-loaded phage syringes against penicillin-resistant *E. coli*, we conducted a series of tests, including DNA quality control, protein quality control, micrograph imaging, and antimicrobial susceptibility testing, with several negative and positive controls. DNA quality control was used to validate the de novo gene synthesis, while protein quality control confirmed cell-free protein expression. Transmission Electron Microscopy (TEM) was employed to examine the morphology, size, and structural integrity of the phage products, as well as to detect potential residues. Prior to antimicrobial testing, the phage products were filtered using size-exclusion chromatography to remove small

molecules and proteins. The antimicrobial activity was then assessed using a standardized suspension of penicillin-resistant *E. coli*.

For negative controls, we used HPLC-grade water (NC) as a blank test, free penicillin G (Pen10ul) as a resistance test, and T7 phage syringes with structure proteins (PSSP) rebooted from structural gene fragments without antibiotic additives as the empty vehicle. Positive controls included T7 phages (PC) rebooted from standard DNA as a natural predator test and T7 phage syringes with all proteins (PSAP) rebooted from all gene fragments without antibiotic additives as a modular genome and *in vitro* expression test.

The antimicrobial susceptibility testing revealed that penicillin-loaded phage syringes exhibit antimicrobial activity against penicillin-resistant *E. coli*. These phage syringes effectively reduced colony-forming units (CFU) of penicillin-resistant *E. coli* compared to the negative controls, including HPLC-grade water (NC), free penicillin G (Pen10ul), and T7 phage syringes with structure proteins (PSSP). However, their effectiveness was lower than that of the positive controls, specifically the standard T7 phages (PC) and T7 phage syringes with all proteins (PSAP). Interestingly, the best-performing phage syringe was not the one expressed with the highest concentration of penicillin G during cell-free expression. Antimicrobial susceptibility testing demonstrated that PS10P was slower and less consistent in reducing the CFU of penicillin-resistant *E. coli* compared to PS1P and PS2P.

This study investigates the potential of phages as delivery vesicles for antibiotic molecules, targeting specific bacterial hosts that would otherwise be impermeable due to antimicrobial resistance mechanisms. We propose that these self-assembling phage syringes, capable of directly injecting antibiotics into bacterial cells, could serve as a novel and specific alternative antimicrobial strategy. Future developments may focus on expanding this proof-of-concept by testing a variety of bacteriophages as delivery vehicles and incorporating different antibiotics as cargo to target a range of multidrug-resistant bacteria. We anticipate that this approach could provide alternative solutions for combating antibiotic-resistant bacterial infections.

## Materials and Methods

### Modular T7 phage genome design and synthesis

We used the NCBI Reference Sequence of Enterobacteria phage T7 (NC_001604.1) to design modular T7 phage genomes. Based on the NCBI RefSeq annotations, the phage genome was identified as structural and non-structural elements. The structural elements were inserted in a linear gene fragment with a T7 promoter sequence and a terminator sequence. The structural gene fragments were capped with universal adapters recommended by the gene synthesis provider (Twist Bioscience, USA). These gene fragments were synthesized into double-stranded DNA via high-throughput silicon-based gene synthesis with an average error rate of 0.013%. The initial oligonucleotides were annealed and PCR amplified on the semiconductor-based silicon platform, followed by error correction through an enzymatic reaction. The non-structural genes were synthesized into double-stranded DNA fragments with the same provider to be used as a positive control of phage syringes. Each synthetic gene fragment was suspended in 5 µL of the TE buffer (10 mM Tris-Cl and 0.5 mM EDTA at pH 9.0) and frozen at -20 °C upon arrival.

### Rebooting T7 phage *in vitro*

For positive control, we obtained a commercially available DNA standard of bacteriophage T7 isolated from an infected *E. coli* strain (Boca Scientific, USA). This DNA standard was used for phage rebooting using the myTXTL Pro Kits (Daicel Arbor Biosciences, USA). The myTXTL kit includes *E. coli* RNA polymerases and T7 RNA polymerases, enabling cell-free protein expression without the need for cloning, cell lysis, or purification steps [51,52]. For a 12 µL myTXTL reaction, 9 µL of Pro Master Mix, 0.5 µL of Pro Helper Plasmid, and 2.3 µL of nuclease-free water were mixed (Table 1). For T7 phage rebooting, a 0.25 nM final genomic DNA concentration was recommended by the manufacturer. The DNA standard of T7 phages was measured to have a concentration of 1126.6 ng/µL (Table 2). To achieve a 0.25 nM genomic DNA concentration in a 12 µL reaction, we added 0.2 µl of the DNA standard of T7 phages to the myTXTL reaction mix in a 1.5 mL tube (Table 1). The myTXTL reaction mix was vortexed briefly and spun down with a mini-centrifuge before being incubated at 27 °C for 16 hours. After the end of the reaction, the myTXTL reaction mix was put on ice.

### Cell-free protein expression of T7 phage syringes

For phage syringes, we first expressed only the synthetic structural gene fragments of T7 phages. The gene fragments of T7 phages were synthesized as described in the previous section. For a 12 µL myTXTL reaction, 9 µL of Pro Master Mix, and 0.5 µL of Pro Helper Plasmid were mixed. For linear DNA templates, a 0.96 nM final genomic DNA concentration was recommended by the manufacturer. The synthetic gene fragments of T7 phages were measured to have a concentration of 143.7 ng/µL (Table 2). To achieve a 0.96 nM genomic DNA concentration in a 12 µL reaction, we added 2.5 µL of the DNA standard of T7 phages to the myTXTL reaction mix in a 1.5 mL tube (Table 1). Additionally, we added 1 µL, 2 µL, and 10 µL of penicillin G to the myTXTL reaction mix for PS1P, PS2P, and PS10P, respectively (Table 1). It was previously demonstrated that the presence of additives such as Glycerol, DMSO, EDTA, Tris-HCL, $CaCl_2$, $MgCl_2$, and NaCl was tolerated without a loss in performance in cell-free protein expression systems [53,54]. The myTXTL reaction mix was vortexed briefly and spun down with a mini-centrifuge before being incubated at 27 °C for 16 hours. After the end of the reaction, the myTXTL reaction mix was put on ice.

For positive control of phage syringes, we also expressed all the synthetic gene fragments of T7 phages. The synthetic gene fragments of T7 phages were measured to have a concentration of 150.6 ng/µL (Table 2). To achieve a 0.96 nM genomic DNA concentration in a 12 µL reaction, we added 2.5 µL of the DNA standard of T7 phages to the myTXTL reaction mix in a 1.5 mL tube (Table 1). The myTXTL reaction mix was vortexed briefly and spun down with a mini-centrifuge before being incubated at 27 °C for 16 hours. After the end of the reaction, the myTXTL reaction mix was put on ice.

## Imaging with Transmission Electron Microscopy

Before imaging, we measured the concentration of the rebooted phage samples with A280 readings from NanoDrop Microvolume UV-Vis Spectrophotometer (Thermo Scientific). We set the molecular weight of A280 readings at the maximum value of 9,999 kPa and extinction coefficient (ε280) at $1,400 \times 10^3 M^{-1} cm^{-1}$. We used 1 µL of each sample to measure A280 and A260 readings to estimate the concentration of phage proteins as well as impurities from cell-free expression (Table 3).

For TEM sample preparation, we fixed each rebooted phage sample on a carbon-coated 400-mesh grid (Ted Pella, USA). A 2 µL of freshly prepared phage sample was applied on the carbon side of the grid for 5 minutes, and the sample was washed with drops of 30 µL of High Performance Liquid Chromatography (HPLC) water (Fisher Scientific, USA). Immediately, the grid was negatively stained with 5 µL of 1% uranyl acetate for 30 seconds. The grid was wiped with Whatmann paper and air-dried overnight before being visualized with TEM. Imaging was performed using Talos F200C G2 at the Imaging and Microscopy Facility of the University of California, Merced.

## Antimicrobial susceptibility testing of controls

For antimicrobial susceptibility testing of controls, we used a standard bacterial suspension of *Escherichia coli* ATCC® 25922™ (Microbiologics, USA). The initial CFU of the standard bacterial suspension was measured with OD600 readings from NanoDrop Microvolume UV-Vis Spectrophotometer (Thermo Scientific). The standard bacterial suspension was incubated at 37 °C for 20 minutes before dosage to ensure the stability of the bacterial cells.

For negative controls of antimicrobial susceptibility testing, we used HPLC water (Fisher Scientific, USA) and penicillin G potassium salt (Fisher Scientific, USA). The first negative control (NC) was conducted by adding 10 µL of HPLC to 600 µL of standard bacterial suspension of *E. coli*. The suspension was incubated at 37 °C and monitored with OD600 readings from the spectrophotometer every 20 minutes for 12 time points after dosage (Table 3). The second negative control (Pen10ul) was conducted by adding 10 μL of penicillin G in distilled and sterile-filtered water (100 mg/mL) to 600 μL of standard bacterial suspension of *E. coli*. The suspension was incubated at 37 °C and monitored with OD600 readings from the spectrophotometer every 20 minutes for 12 time points after dosage (Table 3). Three replicates of the antimicrobial susceptibility test were conducted for each negative control.

      For positive control of antimicrobial susceptibility testing, we used T7 phages rebooted from the standard phage DNA, as outlined in the previous section. The positive control (PC) was conducted by adding 1 μL of rebooted T7 phages to 600 μL of standard bacterial suspension of *E. coli*. The suspension was incubated at 37 °C and monitored with OD600 readings from the spectrophotometer every 20 minutes for 12 time points after dosage (Table 3). Three replicates of the antimicrobial susceptibility test were conducted for the positive control.

## Desalting phage syringes of small molecules

The molecular weight of penicillin G is 334 Da [55], and the residual additives were visible in the TEM micrographs. Before antimicrobial susceptibility testing, the phage syringes expressed *in vitro* using synthetic gene fragments with penicillin as an additive (PS1P, PS2P, and PS10P) were filtered with micro spin desalting columns (Thermo Scientific, USA). We used Zeba Spin Desalting columns designed for desalting proteins with a molecular weight >40 kDA and removing small molecules less than 2 kDa. These micro spin columns contain size-exclusion chromatography resin with a bead structure that can remove low MW contaminants from samples using centrifugal force pressure. The micro spin desalting columns were centrifuged at $1,500 \times g$ for 2 minutes to remove storage buffer and equilibrated with 50 μL of TE buffer by centrifuging at $1,500 \times g$ for 2 minutes. Subsequently, the small proteins and molecules were removed from the phage syringe samples by adding the sample directly onto the resin and centrifuging the micro spin desalting columns at $1,500 \times g$ for 2 minutes. The flow-through containing the sample was retained in a 1.5 mL tube for antimicrobial susceptibility testing.

## Antimicrobial susceptibility testing of phage syringes

For antimicrobial susceptibility testing of phage syringes, we used a standard bacterial suspension of *Escherichia coli* ATCC® 25922™ (Microbiologics, USA). The initial CFU of the standard bacterial suspension was measured with OD600 readings from NanoDrop Microvolume UV-Vis Spectrophotometer (Thermo Scientific). The standard bacterial suspension was incubated at 37 °C for 20 minutes before dosage to ensure the stability of the bacterial cells. The antimicrobial susceptibility test of each phage syringe sample (PS1P, PS2P, PS10P, PSAP, and PSSP) was conducted by adding 1 μL of the sample to 600 μL of standard bacterial suspension of *E. coli*. The suspension was incubated at 37 °C and monitored with OD600 readings from the spectrophotometer every 20 minutes for 12 time points after dosage (Table 3). Three replicates of the antimicrobial susceptibility test were conducted for each sample.

# Declarations

## Ethics approval and consent to participate

Not Applicable

## Consent for publication

Not Applicable

## Availability of data and materials

Supplementary Tables 1-2 (single workbook with multiple tabs). For data analysis, R v.3.6.0 and ggplot2 v.3.3.0 were used.

## Competing interests

H.S. is a founder of BioBCorp, with a patent pending related to this work. The authors declare no competing interests.

## Funding

This research was funded by the startup grant awarded to H.S. at California State University, Fresno. This research received no specific grant from any funding agency.

## Authors' contributions

Experiments were primarily conducted by H.S. Analyses were primarily conducted by H.S. The study was conceived by H.S., and all authors contributed to writing the manuscript.

## Acknowledgments

We thank Kennedy Nguyen at the Imaging and Microscopy Facility at the University of California, Merced for the TEM imaging.

**Figure 1**: Self-assembly of phage syringes with modular genomes via de novo gene synthesis and cell-free protein expression.

Step 1: The bacteriophage genome of interest is designed into modular format. Step 2: The structural elements of the phage genome are synthesized de novo into gene fragments. Step 3: These gene fragments are rebooted as phage syringes in a cell-free protein expression system with varying doses of antibiotics as additives. Step 4: The lytic activities of these phage syringes are validated with antimicrobial susceptibility testing against the host bacteria.

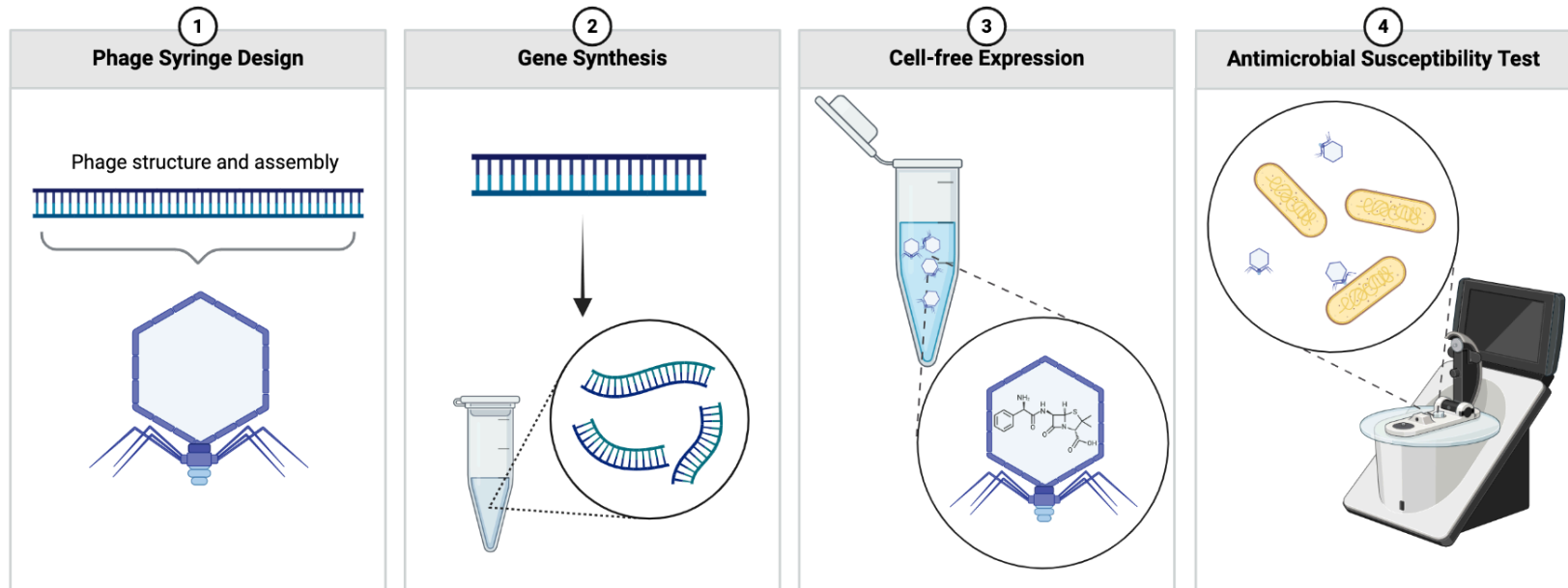

**Figure 2**: The modular genomes of phage syringes to be expressed *in vitro* with penicillin G as additives.

(A) Genomic architecture of T7 phages with lytic activities against *Escherichia coli*. The phage structure and assembly segment is inserted into the gene fragment to be synthesized de novo and expressed *in vitro*. (B) Molecular structure of penicillin G. (C) Phage syringes to be tested for antimicrobial susceptibility in this study include T7 phage syringes assembled with 1 μL, 2 μL, and 10 μL penicillin (PS1P, PS2P, and PS10P, respectively). The positive control is antibiotic-free phage syringes with all proteins (PSAP) and the negative control is antibiotic-free phage syringes with structure proteins (PSSP).

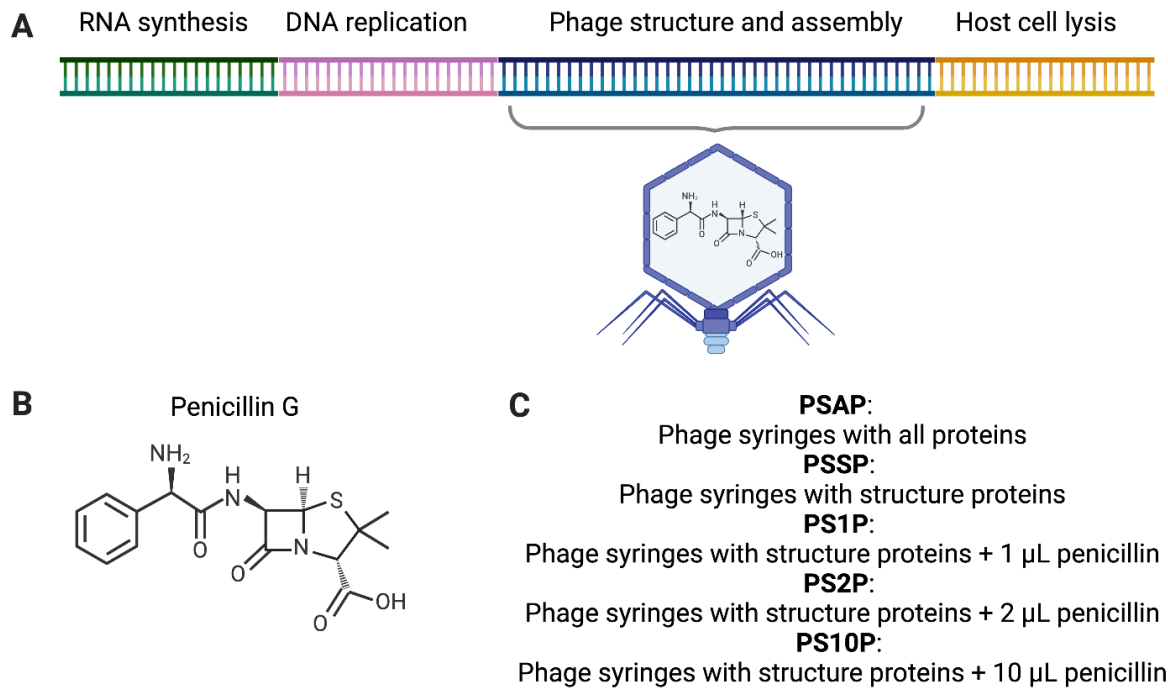

**Figure 3**: Quality control of phage syringes from spectrophotometer readings.

(A) DNA quality control of samples after de novo gene synthesis with 10 mm absorbance readings from NanoDrop Microvolume UV-Vis Spectrophotometer. (B) Protein quality control of samples after cell-free protein expression with 10 mm absorbance from NanoDrop Microvolume UV-Vis Spectrophotometer.

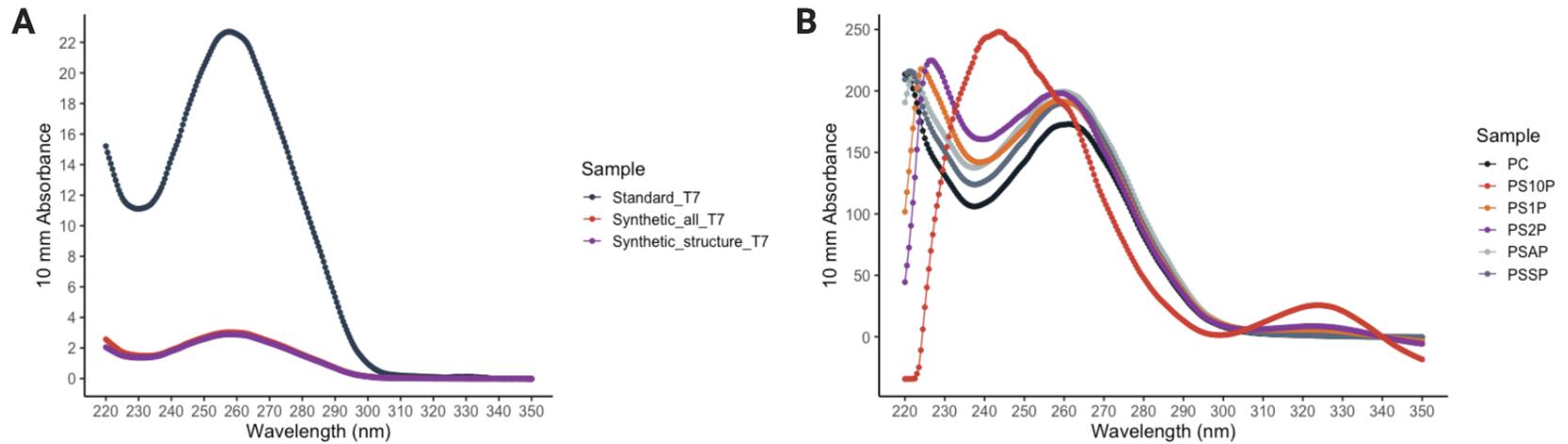

**Figure 4**: TEM images of T7 phages and phage syringes.

Imaging of each rebooted phage sample was performed using Talos F200C G2 fixed on a carbon-coated 400-mesh grid with negative staining. (A) Rebooted standard T7 phages. (B) Rebooted T7 phage syringes with all proteins (PSAP). (C) Rebooted T7 phage syringes with 1 µL penicillin (PS1P).

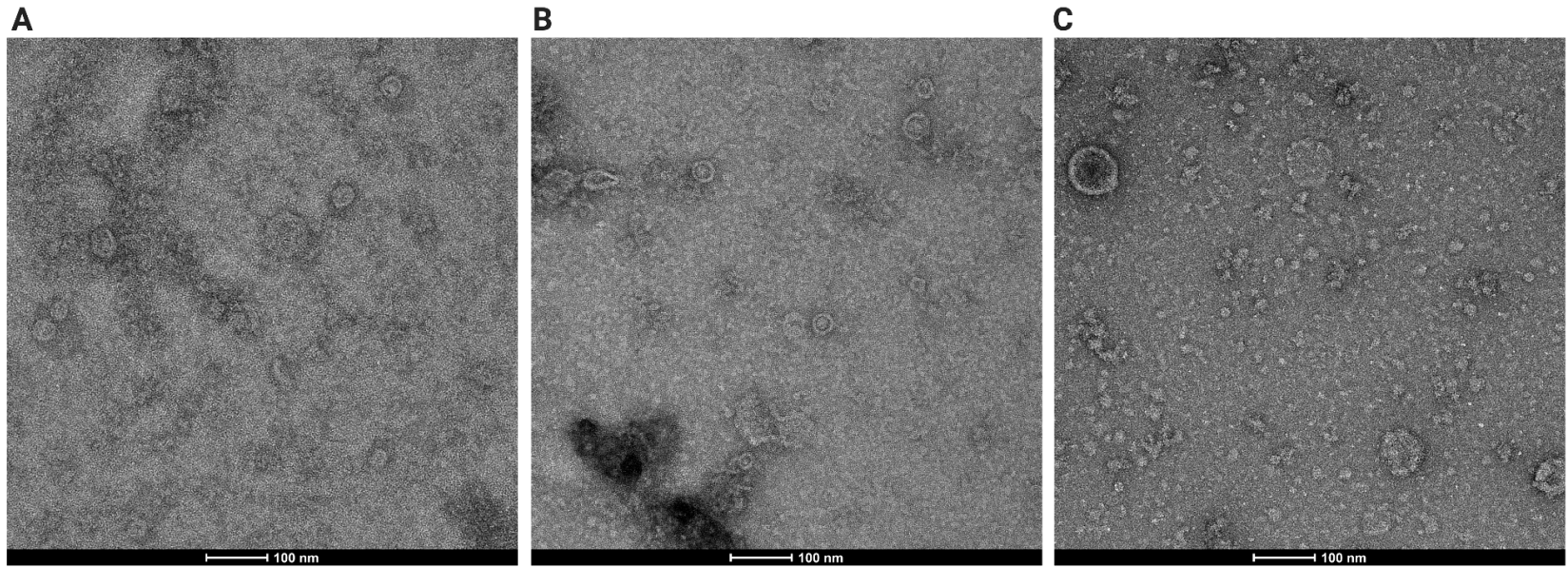

**Figure 5**: Time-serial graphs of antimicrobial susceptibility testing.

Each dot is a normalized OD600 reading from NanoDrop Microvolume UV-Vis Spectrophotometer, measured every 20 minutes for 12 time points after dosage. Three replicates were conducted at each time point for each sample, with the dot showing the average of these replicates. The error bars show the standard deviation between three replicates. (A) Comparison of time-serial OD600 readings of the phage syringes (PS1P, PS2P, PS10P) against those of the negative control using penicillin 10 μL (Pen10ul). (B) Comparison of time-serial OD600 readings of the phage syringes (PS1P, PS2P, PS10P) against those of the positive control using standard T7 phages (PC) and those of the negative control using HPLC water (NC). (C) Comparison of time-serial OD600 readings of the phage syringes (PS1P, PS2P, PS10P) against those of the positive control of phage syringes with all proteins (PSAP) and those of the negative control of phage syringes with structure proteins (PSSP).

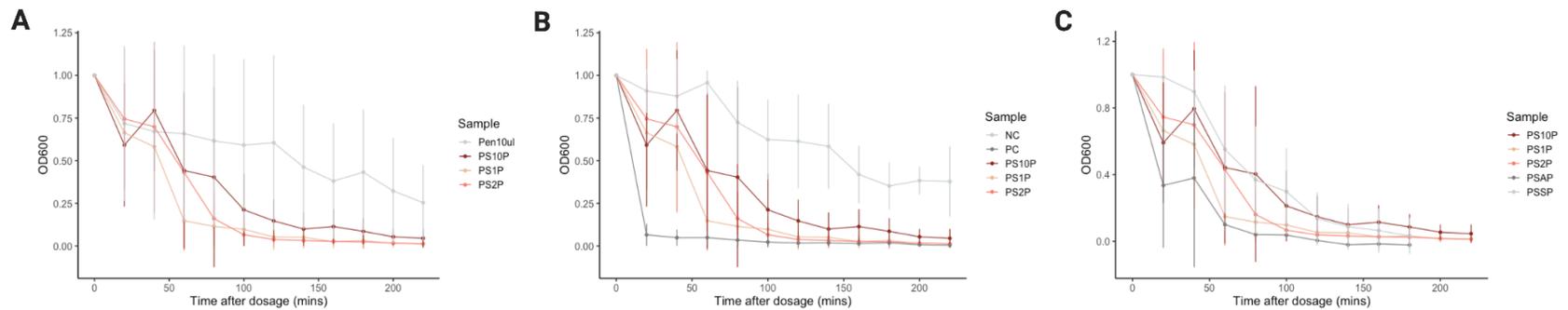

**Table 1**: Experimental parameters for cell-free protein expression of self-assembling T7 phage syringes with modular genomes.

The components and volumes were calculated for a 12 µL cell-free protein expression reaction.

|  | Pro Master Mix | Pro Helper Plasmid | Template DNA | Nuclease-free water | Penicillin G |
|---|---|---|---|---|---|
| **T7 phage rebooting with standard DNA** | 9 µL | 0.5 µL | 0.2 µL | 2.3 µL | - |
| **T7 phage rebooting with synthetic DNA** | 9 µL | 0.5 µL | 2.5 µL | - | - |
| **T7 phage syringes + 1 µL penicillin G** | 9 µL | 0.5 µL | 2.5 µL | - | 1 µL |
| **T7 phage syringes + 2 µL penicillin G** | 9 µL | 0.5 µL | 2.5 µL | - | 2 µL |
| **T7 phage syringes + 10 µL penicillin G** | 9 µL | 0.5 µL | 2.5 µL | - | 10 µL |

**Table 2**: Quality control of DNA and protein during the process of rebooting phage syringes *in vitro*.

(A) DNA quality control of samples after de novo gene synthesis with A260, A230, and A280 readings from NanoDrop Microvolume UV-Vis Spectrophotometer. (B) Protein quality control of samples after cell-free protein expression with A280 and A260 readings from NanoDrop Microvolume UV-Vis Spectrophotometer

| A. DNA quality control | | | | |
|---|---|---|---|---|
| Sample Name | Concentration | Units | A260/A280 | A260/230 |
| Standard DNA of T7 phages | 1126.6 | ng/µL | 1.91 | 2.03 |
| Synthetic all gene fragments of T7 phages | 150.6 | ng/µL | 1.90 | 1.99 |
| Synthetic structure gene fragments of T7 phages | 143.7 | ng/µL | 1.91 | 2.11 |

| B. Protein quality control | | | | |
|---|---|---|---|---|
| Sample Name | Concentration | Units | A280 | A260/A280 |
| Standard T7 phages (PC) | 584.702 | mg/mL | 81.87 | 2.11 |
| Phage syringes all proteins (PSAP) | 697.847 | mg/mL | 97.72 | 2.04 |
| Phage syringes structure proteins only (PSSP) | 644.847 | mg/mL | 90.29 | 2.10 |
| Phage syringes with 1 µL penicillin (PS1P) | 622.620 | mg/mL | 87.18 | 2.19 |
| Phage syringes with 2 µL penicillin (PS2P) | 606.515 | mg/mL | 84.92 | 2.33 |
| Phage syringes with 10 µL penicillin (PS10P) | 341.580 | mg/mL | 47.84 | 3.94 |

**Table 3**: Antimicrobial susceptibility testing of negative controls, positive controls, and phage syringes.

Summary of three replicates of normalized OD600 readings from NanoDrop Microvolume UV-Vis Spectrophotometer every 20 minutes for 12 time points after dosage.

| Time points | PSAP (mean±stdev) | PSSP (mean±stdev) | PS1P (mean±stdev) | PS2P (mean±stdev) | PS10P (mean±stdev) | PC (mean±stdev) | NC (mean±stdev) | Pen10ul (mean±stdev) |
|---|---|---|---|---|---|---|---|---|
| 1 | 1.000±0.000 | 1.000±0.000 | 1.000±0.000 | 1.000±0.000 | 1.000±0.000 | 1.000±0.000 | 1.000±0.000 | 1.000±0.000 |
| 2 | 0.335±0.374 | 0.985±0.026 | 0.664±0.333 | 0.745±0.409 | 0.592±0.361 | 0.066±0.064 | 0.909±0.129 | 0.718±0.454 |
| 3 | 0.379±0.535 | 0.898±0.127 | 0.582±0.388 | 0.699±0.497 | 0.795±0.353 | 0.049±0.048 | 0.878±0.218 | 0.671±0.517 |
| 4 | 0.100±0.036 | 0.552±0.383 | 0.148±0.155 | 0.431±0.458 | 0.443±0.458 | 0.049±0.039 | 0.958±0.071 | 0.659±0.517 |
| 5 | 0.040±0.012 | 0.368±0.321 | 0.115±0.121 | 0.161±0.157 | 0.403±0.527 | 0.035±0.030 | 0.725±0.244 | 0.617±0.508 |
| 6 | 0.037±0.023 | 0.298±0.260 | 0.098±0.079 | 0.066±0.067 | 0.212±0.211 | 0.023±0.031 | 0.624±0.234 | 0.592±0.503 |
| 7 | 0.005±0.002 | 0.137±0.156 | 0.053±0.076 | 0.038±0.044 | 0.147±0.125 | 0.018±0.027 | 0.614±0.274 | 0.605±0.512 |
| 8 | -0.021±0.021 | 0.088±0.137 | 0.051±0.065 | 0.032±0.036 | 0.099±0.097 | 0.018±0.020 | 0.584±0.248 | 0.462±0.365 |
| 9 | -0.016±0.030 | 0.064±0.131 | 0.026±0.039 | 0.027±0.025 | 0.114±0.102 | 0.014±0.016 | 0.419±0.168 | 0.381±0.337 |
| 10 | -0.022±0.023 | 0.032±0.107 | 0.032±0.049 | 0.025±0.036 | 0.086±0.077 | 0.018±0.020 | 0.352±0.137 | 0.433±0.367 |
| 11 | NA | NA | 0.015±0.022 | 0.017±0.017 | 0.054±0.046 | 0.007±0.009 | 0.383±0.081 | 0.323±0.309 |
| 12 | NA | NA | 0.014±0.017 | 0.012±0.015 | 0.045±0.055 | 0.004±0.008 | 0.378±0.205 | 0.254±0.220 |